\documentstyle[12pt]{article}

\newcommand{\beq}{\begin{equation}}
\newcommand{\eeq}{\end{equation}}
\newcommand{\beqa}{\begin{eqnarray}}
\newcommand{\eeqa}{\end{eqnarray}}
\newcommand{\beqar}{\begin{eqnarray*}}
\newcommand{\eeqar}{\end{eqnarray*}}
\newcommand{\tr}{{\rm tr}}

\def \vr {\hat\varrho}
\def \vra {\varrho}
\def \bvr {{\bf\hat\varrho}}
\def \bvra {{\bf\varrho}}
\def \prob {{\cal P}rob}
\def \la {\langle}
\def \ra {\rangle}

\def \up {\uparrow}
\def \down {\downarrow}
\def \o {{\cal O}}
\def \h {{\cal H}}
\def \s {{\cal S}}
\def \md {{\cal MD}}

\begin{document}

\input epsf

 %%%%%%%%%%%%%%%%%%%%%%%%%%%%%%%%%%
\begin{titlepage}

%\vspace{.5in}
\thispagestyle{empty}

\begin{flushright}
TP-94-010\\
TAUP 2200-94\\
%hep-th/yymmddd\\
%October 1994\\
\end{flushright}
%\vspace{0.5in}

\begin{center}
{\bf \Large On a Time Symmetric Formulation of Quantum
Mechanics
}\\

\vspace{.2in}
B. Reznik$^{(a)}$\footnote{\it e-mail:
reznik@physics.ubc.ca} and
Y. Aharonov$^{(b,c)}$\footnote{Also in a visiting
position at Boston University.}\\

\medskip
{(a) \small\it  Department of Physics}\\
{\small \it University of British Columbia}\\
{\small\it 6224 Agricultural Rd. Vancouver, B.C., Canada
V6T 1Z1}\\
\medskip
{(b)  \small\it School of Physics and Astronomy} \\
{\small \it Beverly and Raymond Sackler Faculty of Exact
Sciences}\\
{\small \it Tel Aviv University, Tel-Aviv 69978,
Israel.}\\
\medskip
{(c) \small \it Department of Physics}\\
{\small \it University of south Carolina}\\
{\small\it  Columbia, SC 29208, USA}\\
\end{center}

\vspace{.2in}
\begin{center}
\begin{minipage}{5in}
\begin{center}
{\large\bf Abstract}
\end{center}
{\small
We explore further the  suggestion to describe a pre-
and post-selected system by a two-state,  which is
determined by two conditions.
Starting with a formal definition of a two-state Hilbert
space
and basic operations, we systematically recast the
basics of quantum mechanics - dynamics, observables, and
measurement theory - in terms of two-states as the
elementary quantities.
We find a simple and suggestive formulation, that
``unifies''  two
complementary  observables:  probabilistic observables
and  non-probabilistic `weak' observables.
Probabilities are relevant for  measurements in the
`strong coupling regime'.
They are given by the absolute square of a two-amplitude
(a projection of a two-state).
Non-probabilistic observables are observed in
sufficiently `weak' measurements, and are given by
linear combinations
of the two-amplitude.
As a sub-class they include the  `weak values' of
hermitian operators.
We show that in the intermediate regime, one may observe
a mixing of
probabilities and weak values.

A consequence of the suggested  formalism and
measurement theory,
is that the problem of non-locality and Lorentz
non-covariance, of
the usual prescription with a `reduction', may be
eliminated. We exemplify this point for  the EPR
experiment
and for a system under successive observations.
}
\end{minipage}
\end{center}
\end{titlepage}
%\addtocounter{footnote}{-1}

%%%%%%%%%%%%%%%%%%%%%%%%%%%%%%%%%%%%%%%%%%%%%%%%%%%%%%%%
%%%%%%%%%

\section{Introduction}

Initial and final conditions play  significantly
different
roles in quantum mechanics and classical mechanics.
In classical mechanics the exact state of a system $\s$
at any time  $t$ is determined by a single  condition;
i.e. by feeding the equations of motion with appropriate
initial conditions on a Cauchy surface and working out
the evolution of the system in the future or past.
Traditionally, quantum mechanics is formulated in a
similar manner.
A measurement of a complete set of commuting observables
determines a state $|\psi_1(t_i)\rangle$  of ${\cal S}$;
this provides the initial   condition at $t=t_i$.  To
derive probabilities for various possible measurements
at $t'>t_i$ the Schr\"odinger equation is fed with
$|\psi_1(t_i)\ra$, and  $|\psi_1(t')\ra$ is computed.
Now suppose  we perform at $t=t_f>t'$ another
set of measurements which also determine  the state of
$\s$. While classically, this second measurement is
trivial, in quantum theory the second result
($|\psi_2(t_f)\ra$) is usually not determined from the
initial condition, i.e.,
 in general $|\psi_2(t_f)\ra\ne|\psi_1(t_f)\ra$.
Should we regard $|\psi_2(t_f)\ra$ as a second
condition for the system at  intermediate times
$t_f>t>t_i$?
 After all the  dynamical laws of motion either the
Schr\"odinger or Hiesenberg equations are time
symmetric.
Indeed in quantum mechanics we are free to select
ensembles using two (almost)
independent initial and final  conditions.

In 1964 Aharonov Bergman and Lebowitz \cite{abl} where
the first to recognize the non-triviality of such
circumstances. They have derived the basic expressions
for probability distributions when the physical system
under observation is determined by a pre- and a
post-selection.
More recently the formalism was re-discovered
independently by
Griffiths [2], Unruh [3], and Gell-Mann and Hartle
[4].\footnote{The relation between the approach
developed in this article,  and the
decoherent histories approach is studied elsewhere
\cite{deco}.}
A main elementary observation of these investigations,
which we would like to emphasize,  is that in most
situations, a pre- and post-selected  system {\it can
not be reduced} to an equivalent system with only one
condition, that is, $\s$ can not be described by a wave
function. This observation has been amplified in Ref.
\cite{surprising,spin100}.
It turns out that for rare situations, the outcomes of
{\it ordinary} measurements can yield very strange and
unusual results. It should  be  emphasized however, that
these results are derived by using {\it standard}
quantum mechanics. The `strangeness' of the
results is only due to the  very special  conditions
which where imposed on $\s$.

Nonetheless, the discovery of such new phenomena, was
deeply motivated by a new physical picture, which was
implicitly used already in Ref. \cite{surprising}. In
this picture, the evolution of the wave function in a
pre- and post-selected systems is conceived
in a time symmetric fashion. The two conditions
determine  two wave functions and both are used to
describe the system at intermediate times.  In fact, the
concept of the `weak value' \cite{spin100,weakv},
$A_w$,  of a Hermitian operator $\hat A$, was discovered
while attempting to grasp this  additional information
between two  conditions. In such a weak measurement,
instead of getting one of the eigenvalues of $A$, one
observes a complex number:
$A_w=\la\psi_f|A|\psi_i\ra/\la\psi_f|\psi_i\ra$.
Weak values have been found useful in studying various
problems \cite{brout,englert,steinberg,weakrev}.

However, several basic questions remained. Since in
general the total information on  a pre- and
post-selected system $\s$ can  not be stored in single
wave function,
what is the proper language to describe $\s$ under such
conditions?
In particular, does this mean  that we loss any notion
of a state
at each time slice, or, does
it call for an extension of some of the  basic notions
of quantum mechanics?

Indeed, it has been suggested in Ref.
\cite{albert84I,multiple,complete},
that the usual notion of a state should be generalized.
Generalized states  which are  determined by two
conditions where defined and studied\cite{complete}.
In this work we shall study in more details the
structure and the implications of such a possible
extension.
We shall call the extension of the usual state $\psi$, a
{\it two-state}, and denote it by $\vr$.  Two states,
are elements of an extended Hilbert space which is
equipped  with the   standard set of operations: an
inner product, expansion in terms of basis vectors, and
a projection which yields a two-amplitude. This
two-state Hilbert space is also
further generalized to the case of successive $N$
conditions.

We then systematically recast the basics of  quantum
mechanics -
dynamics, observables, and measurement theory -  in
terms of two-states as the elementary quantities.
What we find is a simple and suggestive formulation that
is particularly suitable to describe systems in a state
of  pre- and post-selection, or  a sub-system which is
coupled to a pre- and post-selected environment
\cite{trace}.
Although our formalism is entirely equivalent to
ordinary quantum mechanics, it
suggests new insights.

Two basic types of observables arise naturally in this
formalism.
In the limit of strong coupling between the measurement
device ($\md$) and $\s$, one measures eigenvalues of
Hermitian operators, but with a probability proportional
to  $|\vra|^2$,
{\it the absolute square of the two-amplitude}, instead
of $|\psi|^2$.
On the other, in the limit of a vanishing interaction
between $\md$ and $\s$,
one generally measures the weak value $A_w$, which is
expressed as a  complex valued linear  combinations,
$\sum a_k \vra_k$, of the two-amplitude $\vra$.
This implies that the weak value should not be given a
probabilistic
interpretation\cite{vaidreality}, but rather should be
understood as
a direct reflection, and hence as a non-demolition {\it
observation of
the two-state amplitude of the system}. In fact, we show
that weak values of Hermitian operators, are only
{\it  a sub-class} of amplitude-like quantities that can
be measured.  For example, we show how the two-amplitude
itself, which is not a weak value of a Hermitian
operator, can still be observed by a suitable weak
measurement.

What happens when the coupling strength between the
observed system and the measuring device is not one of
the latter two limiting cases, but falls in some
intermediate regime? In such a regime, the `reading'
obtained by the measuring device can not be explained in
terms of probabilities nor by weak values
alone.\cite{timegravity}
We shall show that in some cases one measure {\it mixed}
quantities,
which are determined  by probabilities and by weak
values. The observable is then given by an average of
various weak values with a probability distribution of
some set of eigenvalues.

Finally, we argue that our approach has also  some
conceptual advantages.
A major conceptual difficulty in the standard
interpretation is the issue of the `reduction of the
wave function'.
We argue that this  difficulty may be avoided in this
suggested approach. (See also the discussion in
\cite{daniel}). We exemplify this point by showing that
the EPR experiment and the
evolution of a general system under successive
observations, can be described by a two-state without
appealing to a non-local procedure of reduction.
The  non-local collapse is `replaced' by {\it local}
conditions.
The Lorentz
covariance  of our description is obtained by including
the possibility
of correlations between  different times.

The article continues as follows.
In the next section of we
define the basic notion of a two-state Hilbert space and
its further extension to the case of several
conditions, and show  how the two classes of observables
discussed above are expressed in terms of two-states.
In Section 3, we  study measurement theory in terms of
our formalism.
The two limiting cases, of a weak  and a strong
measurement,  are discussed.
We also show that in the intermediate regime,
a new mixing of probabilities and weak values  is
observed.
In Section 4,  we study the  implications to conceptual
problems, such as
the EPR  experiment and to the  situation of successive
observations.
Finally, in the appendix we show how non-generic two-
states, which correspond to  correlation
between initial and final conditions, can be obtained
for an open system.
%\vfill\eject

\section{Time Symmetric Quantum Mechanics}

We start this section by providing the definition of a
two state and constructing  a Hilbert space of
two-states. Then,   we study the basic operations
between two-states and in Section 2.2 we show how to
handle situation with more than two  conditions by using
multiple-states.
The generalized Schr\"odinger
equation for a two-state is presented in Section 2.3,
and in Sections 2.4  and 2.5  we express the
basic observables in terms  of two-states or
multiple-states.

\subsection{Two-States}

Consider a  closed  system $\cal S$ with a given
Hamiltonian $H$, and
two given
conditions, say $|\psi(t_2)\rangle = |\psi_2\rangle$ and
$|\psi(t_1)\rangle
=|\psi_1\rangle$,   $(t_2>t_1)$.
A mild restriction on these  conditions  is
that
\beq
\langle\psi_2|U(t_2-t_1)|\psi_1\rangle\ne0,
\label{restriction}
\eeq
where $U=\exp(-i\int Hdt')$ is the evolution operator,
must be satisfied.
At any intermediate time  $t_2>t>t_1$,  we have both
`retarded' and `advanced' states, $|\psi_1(t)\rangle  =
U(t-t_1)|\psi_1(t_1)
\rangle $ and $|\psi_2(t)\rangle =
U(t_2-t)|\psi_2(t_2)\rangle$, respectively.
We now combine the total information on the state of
$\cal S$ at time $t$, and define a two-state $\vr(t)$ by
\beq
\vr(t) \ \ \equiv \ \ |\psi_1(t)\rangle\langle\psi_2(t)|
{}.
\label{two-state}
\eeq
The two-state is formally an operator and is similar to
the density matrix operator.\footnote
{A closely related object called a `multiple-time state'

was introduced first in  \cite{albert84I,multiple}. The
physical meaning of the two-state we use is  identical
to the `generalized state' defined in  Ref.
\cite{complete}. However, in our notation the two-state
is formally an operator, and therefore simpler to use.}

However, $\vr(t)$ is in general not a Hermitian
operator.
It coincides with the density operator only for two
trivial  conditions.
We shall call a two-state which can be expressed in the
form (\ref{two-state}),
of a direct product of a {\it ket} and a {\it bra},
a {\it generic two-state}.
In the general case, any two-state is an element of a
linear space, $\h_{II}$, of two-states
which we define as follows.
\hfill\break

\vspace{0.3cm}
\noindent
{\bf Definition }\hfill\break
\noindent
{\it Given by a Hilbert  space of states ${\cal
H}_I=\lbrace{|\alpha\ra\rbrace}$, a  linear space of
two-states
${\cal H}_{II}$ is defined
by all the linear combination of generic two-states
$\lbrace|\alpha\rangle\langle\beta|\rbrace$,
where $|\alpha\rangle$ and $|\beta\rangle$
are any two elements of ${\cal H}_I$.}
\vspace{.5cm}

\par
The most general expression for a two-state $\vr\in{\cal
H}_{II}$
is that of a superposition of generic two-states:
\beq
\vr \ \ = \ \ \sum_{\alpha \ \beta}
                 C_{\alpha\beta}\ |\alpha\ra\la\beta| .
\label{gtwo-state}
\eeq
The space ${\cal H}_{II}$ is a Hilbert space with the
{\it inner product} operation  $[15b]$  defined between
$\vr_1, \ \vr_2\in
{\cal H}_{II}$ by
\beq
\la\vr_1,\vr_2\ra \ \equiv \ \tr (\vr_1^\dagger \ \vr_2)
{}.
\label{inner}
\eeq
The trace in Eq. (\ref{inner})  is over a complete set
of states in ${\cal H}_I$.

Due to the restriction   (\ref{restriction}) of
non-orthogonality of the conditions, not all the
two-states in ${\cal H}_{II}$ correspond to physical
states. We define a subspace of physical states,
${\cal H}_{phys}\subset{\cal H}_{II}$, as the collection
of states that satisfy $\tr\vr =\la 1,\vr\ra \not=0$. A
{\it normalized two-state}
will be defined by the condition $\la 1,\vr\ra =1$.

A normalized two-state basis of ${\cal H}_{phys}$ may
then be constructed as follows.
Given by two different orthonormal basis
$S_1=\lbrace|\alpha\rangle\rbrace$ and
$S_2=\lbrace|\beta\rangle\rbrace$ of
 ${\cal H}_I$ with non-orthogonal elements
 ($\langle\alpha|\beta\rangle\ne0, \ \ \forall \ \
|\alpha\ra\in S_1, \  |\beta\ra\in S_2$), the collection
of all the
two-states $\lbrace \vr_{\alpha\beta}\rbrace$ defined by

\beq
\vr_{\alpha\beta}\ \equiv \
{|\alpha\ra\la\beta|\over\la\beta|\alpha\ra}
\ \in \ {\cal H}_{phys},
\label{normalized}
\eeq
forms a normalized two-state basis of ${\cal H}_{phys}$.

Contrary to the usual case, not all the linear
combinations of basis elements
remain in ${\cal H}_{phys}$.
However,  if $dim(\h_{II})=N^2$, then only a $N^2-2$
dimensional hypersurface in this space
is not  in $\h_{phys}$.
Therefore, $\h_{phys}$ is a closed sub-space up to a set
of points of
measure zero.

We also note, that this construction of a normalized
basis is limited to the case of a discrete Hilbert
space. We can use the  basis
$\lbrace{|\alpha\ra\la\beta|\rbrace}$, which has also
the advantage of simplifying  Eq. (\ref{ortho}) and
(\ref{vr1vr2}) bellow, and is somewhat more convenient
for computations. However, as  we shall see in Section
2.4, the  advantage of using the normalized basis
(\ref{normalized}) is that it displays more simply and
directly probabilities in terms of two-states.

The inner product, of two normalized basis elements
satisfies the orthogonality
relation
\beq
\la\vr_{\alpha\beta},\vr_{\alpha'\beta'}\ra = {1\over
|\la\alpha|\beta\ra|^2}
\delta_{\alpha\alpha'}
\delta_{\beta\beta'} .
\label{ortho}
\eeq

Next we define the {\it two-state amplitude}
$\vra(a,b)$, which will play the a role analogue to
$\psi(a)$,
by the projection
\beq
\vra(a,b ) \ \equiv \ {\la\vr_{ab},\vr\ra
\over\la\vr_{ab},\vr_{ab}\ra} \ = \ \langle a
|\vr|b\rangle\la a| b\ra .
\label{amplitude}
\eeq
For example in the case of  a generic normalized
two-state,
$\vr_{12}={|\psi_1\ra\la\psi_2|\over\la\psi_2|
\psi_1\ra}$, the two-amplitude is given by
\beq
\vr_{12}(a \ b )\ = \ {
\psi_2^*(b)\la b|a\ra \psi_1(a)
\over \la \psi_2|\psi_1\ra} .
\eeq
In terms of the two-amplitude, any two-state $\vr$ can
be written as
\beq
\vr \ = \ \int da  db \ \vra(a,b)\vr_{ab} ,
\eeq
and the product between $\vr_1,\vr_2\in{\cal H}_{phys}$
as
\beq
\la\vr_1,\vr_2\ra \ = \ \int
dadb\la\vr_{ab},\vr_{ab}\ra\
 \vra_1^*(a,b)\vra_2(a,b) .
\label{vr1vr2}
\eeq

Note that by simple operations
we obtain a sub-space of ${\cal H}_{phys}$ that can be
mapped back to ${\cal H}_I$.
Given by $\vr\in{\cal H}_{phys}$, say $\vr=
|\psi_1\ra\la\psi_2|$,  we can define an `{\it in }' and
an `{\it out}'
density matrix by
\beq \rho_{in}={\vr\vr^\dagger\over\la\vr,\vr\ra} =
|\psi_1\ra\la\psi_1|
\label{vr1}
\eeq
and
\beq
\rho_{out}={\vr^\dagger\vr\over\la\vr,\vr\ra} =
|\psi_2\ra\la\psi_2|.
\label{vr2}
\eeq
This property can be used to extract from a given
two-state the corresponding  set of  conditions.
However, notice that only in the case that $\vr$ is a
generic two-state,
(i.e. of the form $\vr=|\psi_1\ra\la \psi_2|$) the
conditions
(\ref{vr1}) and (\ref{vr2})
can be represented as pure states. In general,
$\rho_{in}$ and
$\rho_{out}$ have the form  of a mixed states.

Indeed the Hilbert space $\h_{II}$ can be
classify to two basic groups; of generic two-states or
of non-generic two-states, i.e. two-states that can not
be transformed to the generic form.
Generic two-states always satisfy the equation
\beq
\tr(\vr^2) \ \ = (\tr\vr)^2 .
\label{primitive}
\eeq
The physical significance of these  two classes can be
understood as follows. A generic two-state describes a
system $\s$ that is
pre and post selected and possibly observed at  some
intermediate
time by an ``external'' observer as discussed above.
Non-generic two-state, on the other hand, describe
an open system $\s'$, which may be defined by some
division of $\s$ into a sub-system and environment, e.g.
$\s=\s_{environment} + \s'$.
If the total system $\s$ is pre and post-selected but
only observables in $\s'$ are of interested,
then this open system can be described
by a ``reduced''  two-state:  $\vr_{eff}
=\tr_{environment}\vr$.
In general $\vr_{eff}$ is a non-generic two-state.
As is shown in the appendix,  non-generic two-states can
be
obtained even when there is no direct
interaction between the sub-system and the environment.
In this case the
correlations between the system
and the environment are generated by the act of
pre and post selecting measurements.  The
more general case of a direct interaction between the
subsystem and an
environment is discussed in Ref. \cite{trace}.

\subsection{$N$ sequential  conditions and
multiple-states}

In the general case, an arbitrary number of successive
conditions may be
imposed on a single quantum system. These  conditions
may be
independent (up to the restriction of
non-orthogonality), or can be
inherently correlated. Let us impose on the system $N+1$

sequential  conditions
at the times $t=\tau_1,\tau_2....\tau_{N+1}$. We have
already constructed a Hilbert
space of two-states for the case of only two
conditions. Let us consider  only such two sequential
conditions, at  $\tau_i$ and $\tau_{i+1}$, and  for a
moment  ignore all the other conditions. At this $i$'th
time interval, we can construct as before a two-state
$\vr^{(i)}(t)$, where $t_i\in(\tau_i,\tau_{i+1})$, which

is an element of the Hilbert space ${\cal
H}_{phys}^{(i)}$ defined above.

A  `generic' multiple-state $\bvr_{abc...z}$ that
describes the system
in the interval $t\in(t_1,t_N)$ is defined as an element
of a  Hilbert space
formed by the direct product
\beq
\bvr_{abc...z} \ \in \ {\cal H}_{phys}^{(1)}\otimes{\cal
H}_{phys}^{(2)}\otimes
  \cdot\cdot\cdot\otimes{\cal H}_{phys}^{(N)}
\eeq
or expressed in terms of normalized  basis elements:
\beq
\bvr_{abc...z}(t_1,t_2,...,t_N)\ = \
\vr^{(1)}_{ab}(t_1) \otimes
\vr^{(2)}_{bc}(t_2)
 \otimes \cdot\cdot\cdot\otimes
\vr^{(N)}_{yz}(t_N) .
\label{generic-multiple-state}
\eeq
The most general multiple-state may also describe
correlations between
various  conditions. Therefore, in general
\beq
\bvr(t_1,t_2,...,t_N) \ = \ \sum_{abc...z}C_{abc...z}
\bvr_{abc...z}(t_1,t_2,...,t_N) .
\label{multiple-state}
\eeq
Therefore, in the case of $N+1$ conditions, the most
general multiple state is an element of the Hilbert
space
which is defined by:  $\h_{N+1} =
\lbrace{\bvr_{abc...z}\rbrace}$, i.e. by all the linear
combinations of generic multiple states.
When the  conditions are not correlated, as in
the case of $N+1$ independent measurements, the
expression for
the multiple state $\bvr$ has the form of the  generic
state in
(\ref{generic-multiple-state}).

The generalizations of the inner product and of the
projection of the
multiple-state to a multiple-amplitudes  are
straightforward.
The inner product
between generic multiple-states is generalized to
\beq
\la\bvr_{a,b,c...,z},\bvr_{a',b',c'...,z'}\ra \ = \
{1\over|\la a|b\ra\la b|c\ra\cdot\cdot \cdot\la
y|z\ra|^2}
\delta_{aa'}\delta_{bb'}\delta_{cc'} \cdot\cdot\cdot
\delta_{zz'}
\eeq
and for any to multiple states
\beq
\la\bvr_1,\bvr_2\ra \ = \
\sum_{aa'bb'cc'...zz'}C^*_{1abc...z}C_{2a'b'c'...z'}
\la \bvr_{abc...z},\bvr_{a'b'c'...z'} \ra .
\eeq
We define the  multiple-state  amplitude according to
equation
(\ref{amplitude}) as
\beq
\bvra(a,b,c,...,z;\ t_1,t_2,...,t_N) \ = \
{
\la \bvr_{abc...z},\bvr(t_1,t_2,...,t_N) \ra
\over
\la  \bvr_{abc...z},\bvr_{abc...z} \ra } .
\eeq
When the multiple-amplitude is expended in term of the
normalized basis,
 the expansion coefficients are given by the
multiple-amplitude:
\beq
\bvr(t_1,t_2,...,t_N) \ = \ \int
dadb...dz\bvra(a,b,...,z;\ t_1,t_2,...,t_N)
\bvr_{abc...z}(t_1,t_2,...,t_N) .
\label{multis}
\eeq
The inner product generalizes to
\beq
\la \bvr_1,\bvr_2 \ra \ =  \
\int da db ...dz \la \bvr_{ab...z},\bvr_{ab...z} \ra
\bvra^*_1(a,b,...,z)\bvra_2(a,b,...,z) .
\eeq

As in the case of two-states, multiple states also be
classified according to Eq. (\ref{primitive})
to generic and non-generic states.
The latter case   corresponds to correlations between
the  conditions at various times.

\subsection{Dynamics}

Two states satisfy the  Liouville equation
\beq
i\hbar \partial_t \vr(t) \ \ = \ \ \lbrack H,\vr(t)
\rbrack  .
\label{motion}
\eeq
Expanding in terms of the two-amplitude we can obtain a
Schr\"odinger-like equation.
For example, if $H=\hat p^2/2m+V(\hat x)$, the
two-amplitude in the coordinate basis,
$\vra(x',x'',t)=\la   x'|\vr(t)|x''
\ra$, satisfies the  equation
\beq
i\hbar \partial_t\varrho(x',x'',t) \  = \
-{\hbar^2\over2m}\biggl(\partial_{x'} -
\partial_{x''}\biggr)
\varrho(x',x'',t) +
\biggl( V(x') - V(x'') \biggr)\varrho(x',x'',t)
\label{schrodinger-like}
\eeq
$$
=\Bigl( H(x',p') - H(x'',p'')\Bigr)\vra(x',x'',t) .
$$
The evolution operator is therefore given by
\beq
U(t) = \exp\Biggl\lbrace
-{i\over\hbar}\int dt\Bigl(H(x',p')- H(x'',
p'')\Bigr)\Biggr
\rbrace .
\eeq

Clearly, for any solution of  (\ref{motion}) or
(\ref{schrodinger-like})
we can construct appropriate conditions, and vice versa.
We also note that the scalar product $\la\vr_1,\vr_2\ra$

is conserved under the
evolution.
Therefore $U$ is a unitary operator in the Hilbert space
$\h
_{phys}$.

{}From (\ref{schrodinger-like}) we can derive the
(generalized) continuity equation
\beq
\partial_t(\vra_1^*\vra_2) + \partial_{x'}{\cal J}' -
\partial_{x''}{\cal J}'' = 0 ,
\label{continuity}
\eeq
where the two-current ${\cal J}'$ is given by
\beq
{\cal J}'(x'\ x''\ t)
 \ = \ {\hbar\over2im}\biggl(\varrho_1^*(x'\ x''
t)\partial_{x'}
\varrho_2(x'\ x''\ t) - c.c.\biggr) ,
\eeq
and ${\cal J}''$ by a corresponding equation.

To get the equation of motion for the  multiple-state
case, we simply need to replace (\ref{motion}) by an
$N$-times generalization:
\beq
i\hbar\Bigl(\partial_{t_1}+\partial_{t_2}+...+\partial_{
t_N}\Bigr)
\vra(t_1,t_2,...,t_N)=
[ H, \vra(t_1,t_2,...,t_N) ].
\label{ms}
\eeq
The multiple-states defined in Section 2.2 are solutions
of (\ref{ms}) and are determined by $N+1$  conditions.

\subsection{Probabilistic observables}

Given an ensemble of $n$ different particles, all in the
same two-state,
we may perform a measurement of an observable $A$.
To this end,  $n$ different
measurement devices
are couple to each of the components of the two-state of
the ensemble
\beq
\vr_{ensemble} =
\vr(1)\otimes\vr(2)\otimes\cdot\cdot\cdot\otimes\vr(n).
\eeq
Each of the measurements will yield as an outcome one of
the eigenvalues
$a$ of the Hermitian operator $A$ with a probability
$\prob(a)$.
This probability was evaluated first in Ref. \cite{abl}.
 In our notation we find
\beq
\prob(a) \ = \ {|\tr(\vr_{aa}\vr)|^2\over\int da
|\tr(\vr_{aa}\vr)|^2} =
 {|\la\vr_{aa},\vr\ra|^2\over\int
da|\la\vr_{aa},\vr\ra|^2} ,
\label{prob1}
\eeq
or in terms of the two-amplitude $\varrho(a,a)$
\beq
\prob(a) \ = \ {|\varrho(a,a)|^2\over\int
da|\varrho(a,a)|^2 } .
\label{proba}
\eeq
The last expression for the probability is of particular
interest. We see
that the projection of the two-state $\varrho(a,a)$
behaves as an
amplitude. The absolute square of the two-amplitude
yields the probability.
The expression for the average value of the observable
$A$ is simply
\beq
\langle A\rangle \ = \
{\int da \ a|\varrho(a,a)|^2\over\int da|\varrho(a,a)|^2
} .
\label{strong-value}
\eeq

Does $\vra(a,b)$, the non-diagonal element of the
two-state,
 correspond to a physical amplitude?
Remember that the two-state $\vr$ may be written as a
linear superposition
of two-states $\vr_{ab}$ with a (complex) amplitude
$\vra(a,b)$:
\beq
\vr \ = \ \int da db\ \vra(a,b)\vr_{ab}
\eeq
A straightforward computation confirms that the absolute
square
of $\vra(a,b)$ yields the probability to find the
generic
two-state $\vr_{ab}$. In other words, if we would
measure
 first the operator $A$ at time $t$
and then the operator $B$ at time $t+\epsilon$,  then
(when $\epsilon\to0$) the probability to find he
eigenvalues $a$
and $b$ is given by
\beq
\prob(a,b) = {|\vra(a,b)|^2\over\int dadb\
|\vra(a,b)|^2}
\label{probab}
\eeq
Equation (\ref{proba}) above corresponds to the special
case of a two-state $\vr_{ab}=\vr_{aa}$.

Comparing to the ordinary expressions when only a
pre-selection is involved, we notice that the
normalization  $\int dadb\ |\vra(a,b)|^2$  above, or in
Eq. (\ref{proba}), is not a constant of motion. It is
also interesting to note that the two-amplitude is
generally a product of two wave functions.
For example,  if $\psi_1(x)$ is pre-selected and later
$\psi_2(x)$ is post-selected, then the (non-normalized)
two-amplitude in this case is
\beq
\vra(x,x,t)=
\psi_2^*(x)U^\dagger(t-t_2)U(t-t_1)\psi_1(x)
\label{vraxx}
\eeq
It is amusing, that when $H=0$,  and the same state is
pre- and post-selected,  the two-amplitude
$\vra=|\psi|^2$ plays also the role of a measurable
probability. In the next section we shall see that this
probability
can also be re-written as a weak value.

All the expressions above are generalized directly to
the case of
a multiple-state. Given by an ensemble of system with
the same
multiple-state, we can measure various Hermitian
operators
at any of the $N$ time intervals.
Let us denote these operators by
$A^{(1)},B^{(2)},...,Z^{(N)}$ and their
eigenvalues by $a,b,...,z$.
The latter operators act on elements of the two-state
Hilbert spaces
${\cal H}_{phys}^{(1)},{\cal
H}_{phys}^{(2)},...,\h_{phys}^{(N)}$, respectively.
The probability to obtain  the values $a,b,c,...,z$
for $N$ measurements, one at  each interval,
is given by
\beq
\prob(a^{(1)},b^{(2)},...,z^{(N)}) \ = \
{  |\bvra(a,a,b,b,...,z,z)|^2
\over
\int da'db'...dz' |\bvra(a',a',b',b',...,z'z')|^2 } .
\eeq
When two measurement are performed at each interval, say
$A^{(1)}$ and $B^{(1)}$
on the first interval etc., we find
\beq
\prob(a^{(1)},b^{(1)},...,y^{(N)},z^{(N)}) \ = \
{  |\bvra(a,b,...,y,z|^2
\over
\int da'db'...dz |\bvra(a',b',...,y',z'|^2 } .
\label{probgen}
\eeq

Therefore, the coefficients in the expansion
of the multiple-state in  (\ref{multis}) correspond,  in
this general case as well,
to physical amplitudes.

Having spelled out the general expressions, we can
easily verify that
they are time symmetric.
Taking $t\to -t$, corresponds to the transformation $\vr
\to \vr^\dagger$
or to replacing  the two-amplitude $\vra$ by $\vra^*$.
Clearly this transformation does not affect  Eq.
(\ref{probab}) or (\ref{probgen}).

Finally,  we would like to show that all the usual
probabilistic
information in the case of an ensemble with only one
condition is contained in our formalism. Given by two
conditions,
say $|\psi(T)\ra=|\psi_2\ra$ and
$|\psi((-T)\ra=|\psi_1\ra$,
the two-state $\vr$ is determined. But now suppose we
are given by
$\vr$ and we would like to reconstruct
the probabilistic quantities related to an ensemble
which is only pre (post) -selected, i.e.  with
only one given condition   $|\psi_1\ra$ ($|\psi_2\ra$).
In this case
the probability $\prob_I(a)$  to measure
the state $|a\ra$ is given simply by
\beq
\prob_I(a) \ = \ |\la a|\psi_1\ra|^2 \ = \
 {           \la \vr_{aa},\rho_{in}\ra },
\label{probI}
\eeq
(or by $ \la \vr_{aa},\rho_{out}\ra $),
where $\rho_{in}$ and $\rho_{out}$ where defined in
(\ref{vr1},\ref{vr2}).
(In fact, as shown in Section 4.2, Eq. (\ref{probI}) can
be reconstructed
directly from Eq. (\ref{prob1}).)
The expectation value of an hermitian operator for a
pre-selected ensemble is simply given by
\beq
\langle A \rangle_I =\tr A\rho_{in}=
{\langle \vr, A\vr \rangle \over \langle\vr,\vr\rangle
}.
\eeq
Viewing the two  conditions   as results of measurements

we can also ask what is the probability to get
$|\psi_2\ra$ given by an
ensemble described by $|\psi_1\ra$. This probability is
given by
\beq
\prob_I(\psi_1\to\psi_2) \ = \ |\la\psi_2|\psi_1\ra |^2
\ = \  \la \rho_{out},\rho_{in}\ra.
\label{prob12}
\eeq

\subsection{Non-probabilistic observables and `weak
values'}

Given by a pre- and post- selected
ensemble the weak value of an operator $\hat A$ is
defined \cite{spin100} by
\beq
A_w \ = \ {\langle\psi_2|A|\psi_1\rangle
           \over
          \langle\psi_2|\psi_1\rangle } .
\label{old-weak}
\eeq
The weak value is in general a complex quantity.
However,
{\it both} the real
and the imaginary parts of the weak value are observable
quantities\cite{spin100} (and see Section 3.4).
We shall argue that the weak values are only a subclass
of the
non-probabilistic observables that are available to us.

Let us see how observables of the weak type are
expressed in our notation.
Given by a two-state $\vr$,  Equation (\ref{old-weak})
can be written as\footnote{A similar expression for weak
values was found also in Ref. \cite{complete}. }
\beq
A_w \ = \
          { \tr(A\vr) \over \tr\vr }
      =   { \la A,\vr\ra \over \la 1,\vr\ra}
\label{weak}
\eeq
or in terms of the two-amplitude $\vra(a,a)$ we have
\beq
A_w \ = \ {  \int da \ a \varrho(a,a)
             \over
             \int da \varrho(a,a) }.
\label{weak-amplitude}
\eeq
This expression is correct also for the more general
case of non-generic two-states

The last expression for the weak value is of particular
interest. Comparing
this equation to expression (\ref{strong-value}) for the
expectation
value of operator, we note that the weak value is given
by an {\it average} of
a two-amplitude rather then  the square of the
absolute value of a two-amplitude.
The weak value is in fact a measure of the two-amplitude
itself. Inserting for
$A$ a projection operator $\pi_a=\vr_{aa}$, we get
\beq
(\pi_a)_w=(\vr_{aa})_w \ = \ \varrho(a,a).
\eeq
Therefore the weak value of a Hermitian operator is
simply
a superposition of the diagonal elements of the
two-amplitude.

We now see that there is no basic difference between the
physical
interpretation that should given to the weak value of a
Hermitian operator and
to the components of a two-state. In fact the
two-amplitude,
say $\vra(a,b)$, can also be represented
as a weak value of the non-Hermitian operator
(two-state) $\vr_{ab}$
\beq
\vra(a,b) = {(\vr_{ab})_w \over\la\vr_{ab},\vr_{ab}\ra}
{}.
\eeq
We shall see in the next section that although
$\vra(a,b)$ corresponds to the weak value of a
non-Hermitian observable it can still be  measured.

As a consequence of Eq. (\ref{weak}) the weak
observables share the linearity
property of two-states.
Given by the two-states $\vr_1$ and $\vr_2$ we may
construct
by superposition the two-state $\vr = c_1\vr_1 +
c_2\vr_2$.
The weak values of an observable $A$
satisfies the same linear relation
\beq
A_w(\vr) \ = \ c_1A_w(\vr_1) \ + \ c_2A_w(\vr_2).
\label{superposition}
\eeq
Here $A_w(\vr)$ stands for the weak value of an
observable $\hat A$
for a system with a two-state $\vr$.
This additivity of weak values can now be understood as
a natural
 consequence of a {\it superposition principle} for
two-states, or two-amplitudes.

Equation (\ref{superposition}) can be further
generalized.
Given by the weak value of an operator $A$ with respect
to the two-state $\vr$
we can express this weak value with respect to an
arbitrary basis,
${\vr_{ab}}$ of ${\cal H}_{phys}$, by the transformation
law
\beq
A_w(\vr) \ = \ \int dadb \ \vra(a,b) A_w(\vr_{ab})
\label{transformation}
\eeq
Notice that this is exactly the same expression for
decomposing a two-state
$\vr$ in term of the basis $\vr_{ab}$.
Hence, Equation (\ref{transformation}) expresses an
interesting inner-relation between probabilistic
and non-probabilistic quantities. If we could measure
strongly $\vr_{ab}$
and simultaneously the weak value of $A$ in the `branch'
$\vr_{ab}$ of $\vr$, we would obtain the value
$A_w(\vr_{ab})$ with a probability given by the square
of the two-amplitude!
It is amusing that such a circumstances does in fact
occur, for  measurements of intermediate coupling
strength. This will be further discussed in  Section
3.3.
\vfill\eject

\section{Time Symmetric Description of Measurements}

In this section we shall examine the relation between
the two classes of observables,
which were defined in the last section, to
measurements. We first give a
time symmetric description of a measurement in a pre-
and post- selected
ensemble.

Consider a system $\cal S$
with a given Hamiltonian $H_{\cal S}(x,p)$ and a
measuring device $\cal MD$
with a Hamiltonian $H_{\cal MD}(q,\pi)$.
The measurement process of an observable $A(x,p)$ is
described by
 coupling $\cal S$ and
$\cal MD$ via and some interaction term
$H_{I}$. The prescription of von-Neumann is to take
\beq
H_I \ = \ g(t)qA
\label{md}
\eeq
and use the canonical variable $\pi$ as the `pointer' of
the measuring
device. For $g(t)=g_0\delta(t)$, the shift in the
pointer's location is $\delta\pi=\pi_f-\pi_i = g_0A$.
In this impulsive limit, the free part of $H$ has no
effect. Therefore,
for simplicity we shall set in the following $H_{\cal
MD}=H_{\cal S} = 0$.

The Hilbert space of the total system
is ${\cal H}={\cal H}_{\cal S}\otimes{\cal H}_{\cal
MD}$.
Given by two (consistent)
 conditions, say $\vr(-T)\vr^\dagger(-T)=\rho_1
=|\psi_1\ra\la\psi_1|$ and
$\vr^\dagger(+T)\vr(+T)=\rho_2=|\psi_2\ra\la
\psi_2|$,  we now wish to solve equation (\ref{motion})
and find
$\vr(t)$ in the time interval $t\in[
-T,+T]$. The consistency of the two  conditions is that
our
solution must satisfy $\tr\vr\not=0$, or $\la \rho_1,
\rho_2\ra\ne0$,
which meaning that there is  a finite
amplitude for the system to evolve the initial  to the
final condition.

The Schr\"odinger equation for the (non-normalized)
two-amplitude,
  $\rho(a,a',\pi,\pi',t)=\la a,\pi|\vr(t)|a',\pi'\ra$, is
\beq
i\hbar \partial_t \rho(a,a',\pi,\pi',t) \ = \
-ig(t)\Bigl(a{\partial\over\partial\pi} -
a'{\partial\over\partial\pi}\Bigr)\rho(a,a',\pi,\pi',t).
\label{meas}
\eeq
The two-amplitude may be decomposed as
$\rho=\psi_1(a,q,t)\psi_2(a',q',t)$
where $\psi_1$ and $\psi_2$ are the ordinary wave
functions with
Hamiltonians $H(a,q,t)$ and $-H(a',q',t)$, respectively.

The two-state is therefore given by
\beq
\vr(t) = |\psi_1(t)\ra\la\psi_2(t)|
\eeq
with $|\psi_1(t)\ra=U(t+T)|\psi_1\ra$ and
$|\psi_2(t)\ra=U(t-T)|\psi_2\ra$.

\subsection{Measurements with a probabilistic outcome.}

Consider a measurement of an observable $\hat A$ with
discrete eigenvalues which for simplicity we set to be:
$a=0,\pm1,,,,\pm n..$.
In the idealized  description (\ref{md}) of a
measurement given above, the accuracy in reading $A$ is
given by $\Delta A = \Delta\pi/g_0$, where $\Delta\pi$
is the uncertainty in the initial and final locations of
the pointer, i.e. $\Delta\pi\simeq
\Delta\pi_i\simeq\Delta\pi_f$.
Remembering that the spectrum of $A$ is discrete with
intervals of $1$, we can now say that for an accurate
measurement  we must set
\beq
{\Delta\pi\over g_0 } <<1
\label{strongc}
\eeq
We now notice that, this conditions also implies that
the uncertainty in the interaction term must be very
large, that is, $\Delta(H_I) =(g_0/\Delta\pi)A>>A$. We
shall call this type of
measurements, {\it  strong measurements}, since while
the value of $A$ is unchanged
($[A,H_I]=0$) any other quantity which does not commute
with $A$
is disturbed strongly.  This of course reflects the
consistency of measurement theory with the uncertainty
principle. In the next section we shall see what happens
if one tries to relax Eq. (\ref{strongc}).

Let us consider as an
example, a measurement of $A$ with an outcome
$\delta\pi=\pi_f-\pi_i=1$. The measuring device was
prepared at the state
$|\pi(-T)=0\ra$ and was determined in the final state to
be in the state $|\pi(+T)=1\ra$.
Let us also assume that the initial and final states
 of the observed system were
$|\psi_1(-T)\ra=\sum_nC_n|n\ra$ and
$|\psi_2(T)\ra=\sum_mC'_m
|m\ra$, respectively. This is a complete specification
of two  conditions for
the total system.
The  interaction (\ref{md}) between the measuring device
and the system
occurred at the instant $t=0$ and for the rest of the
interval there is no
evolution, $H_{total}=0$. Therefore, we can easily
derive the two-state
of the total system.
\beq
\bvr(t)=N\sum_{nm}C_nC'^*_m\Bigl(|\pi=n\ra\la\pi'=1|\Bigr)\otimes
\Bigl(|n\ra\la m|\Bigr), \ \ \ \ t\in(0,+T)
\label{mdsone}
\eeq
and
\beq
\bvr(t)=N\sum_{nm}C_nC'^*_m\Bigl(|\pi=0\ra\la\pi'=1-m|\Bigr)\otimes
\Bigl(|n\ra\la m|\Bigr), \ \ \ \ t\in(-T,0)
\label{mdstwo}
\eeq

A schematic description of the evolution of
the wave functions due to the measurement is depicted in
Figure  1.
In the `forward' time direction (upwards in Fig. 1), the
single component $\pi=0$
of the measurement device `splits' at $t=0$ to  discrete
branches
according to the possible final values of $\pi$.
 The forward moving (retarded) state is a product state,
$|\pi=0\ra
\otimes\sum_aC_a|a\ra$,
before the instant of interaction, and an entangled
state,
$\sum_nC_n|\pi=n\ra\otimes|n\ra$ for $t\in(0,+T)$
(correlated states are depicted by doted arrows).
The backwards moving wave behaves symmetrically.
 The advanced state
is given by a direct product for $t\in(0,+T)$, and by an
entangled state
for $t\in(-T,0)$. The two-state of the system
(\ref{mdsone})
is a product of the corresponding
forward (retarded) state, and backwards (advanced)
state.
\medskip
\begin{center}
\leavevmode
\epsfysize=7.0cm
\epsfbox{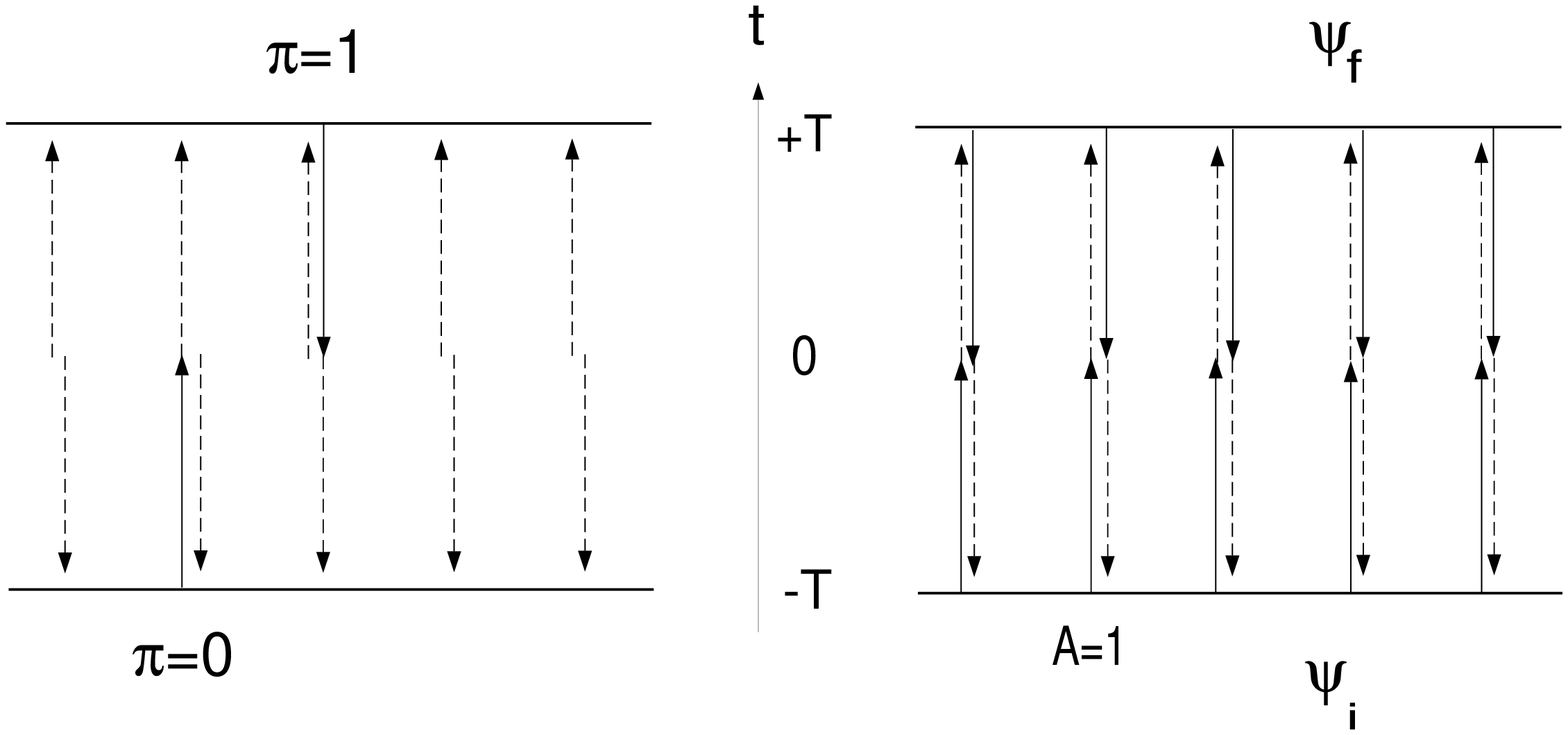}
\end{center}
\noindent
{\small Figure 1.
A pictorial description of the two-state $\vr(t)$  give in
 eq.
(\ref{mdsone}) and  (\ref{mdstwo})  of a
measuring device $\md$
and a system $\s$ during a measurement, in
the special
case that the result  $\pi_f-\pi_i=A=1$ was recorded.
The system and the measuring device are pre-selected
to the state $|\psi_i\ra$ and $|\pi_i=0 \ra$ at $t=-T$,
and post selected to $|\psi_f\ra$
and $|\pi_f=1\ra$ at $t=+T$. The interaction between
$\md$ and $\s$ occurs at $t=0$.
Time flows in the upwards direction, while the
horizontal axis describes the internal
space of $\md$  (left) and $\s$ (right).
Arrows in the up (down) direction represent  ``ket''
(``bra'') components of $\vr$ that
evolve forward (backward) in time. E.g. for
$t\in(-T,0)$,
in the forward time direction,  $\vr$ has  only
one  component of $\md$ with $\pi=0$. After the
interaction,  for $t\in (0,T)$ the two-state $\vr(t)$
has several components of $\md$
that propagate  forward in time. These states are
entangled with
forward evolving states  of $\s$.
Whenever, such entanglement occurs we use dashed lines.
Undashed lines represent the case of a direct product.
}.

\medskip

How can we extract the ordinary (only pre-selected)
 probabilities  from this
picture?
Clearly given by only one  pre- and post-selected
ensemble we cannot.
However, we can consider different ensembles  and
compute the conditional probability
to find $\pi'=1$ when $\pi=0$ and the initial
and final states of $\cal S$ are given. This yields:
\beq
\prob(\pi'=1) \ = \
{\prob_I(\pi=0\to\pi'=1|\psi_i({\cal S}),\psi_f({\cal S}))
\over
\sum_{\pi'=n}\prob_I(\pi=0\to\pi'|\psi_i({\cal S}),
\psi_f({\cal S}))}
\eeq
Using Eq. (\ref{vr1},\ref{vr2},\ref{prob12}) we get
\beq
\prob(\pi'=1) \ = \
{\la \rho_{out}(\pi'=1),\rho_{in} \ra \over
\sum_{\pi'=n}\la \rho_{out}(\pi'),\rho_{in} \ra} \ = \
{|C_1C'_1|^2\over\sum_n |C_nC'_n|^2 },
\label{same}
\eeq
which is of course identical to the probability derived
in this case from
Eq. (\ref{proba}).

We now observe that in the two-state formulation we do
not need to invoke any assumption on a non-local
reduction
of the wave function of ${\cal S}$ due to the (final)
determination of the
measuring device. The traditional  formulation
of the measurement process states that after determining
the location of
the pointer the wave function of the pointer {\it and of
the system}
are reduced instantly to one of the components
$|\pi=1\ra|A=a\ra$. This reduction, is frequently a
non-local process. For example,
we could make the final measurement of the location of
the pointer
(coupling to a external macroscopic environment)
after separating ${\cal S}$
and ${\cal MD}$ to a large distance from each other.
Contrary to the usual description in this symmetric
formulation
of quantum mechanics we need to
invoked only two {\it local}  conditions
on the system and the measuring device to fully
determine the two-state. Thus the determination of the
final location of the pointer reduces only the location
of the pointer,  but does not affect (via a collapse)
the system.

To exemplify this point let us return to the measurement
above but
view the process in two different Lorentz frames ${\cal
O}_1$ and ${\cal O}_2$
with velocities $\vec v_1=v\hat x$ and $\vec v_2 = -v
\hat x$, respectively.
To make the argument clearer let us assume that the
measurement process described above
takes place in the following way.
$\md$ and $\s$ are post selected (prepared)
at $t=-T$ at two different locations, say $x_{\md}=-L$
and $x_\s=+L$.
$\md$ and $\s$ are then transported to one location, say
$x=0$, and interact at $t=0$ via a von-Neumann coupling
(\ref{md}).
They are then transported back to $x_\md$ and $x_s$, and
at $t=+T$ they are
post selected, i.e. coupled to a macroscopic device that
determines the
final states $|\pi_f\ra$ and $|\psi_f\ra$ of $\md$ and
$\s$, respectively.
We assume that the variables $\pi$ and $A$
are  internal local degrees of freedom. Therefore the
process of pre and post
selection and the interaction can taken as local. In the
original (stationary)
frame the evolution in this internal space is depicted
in Fig. 1.

Clearly,  as the preparation (or post selection ) of
$\md$ and $\s$ take place in space-like separated
locations,
the temporal order of the events is
different in  ${\cal O}_1$ and ${\cal O}_2$. In  ${\cal
O}_1$, an
observer sees the post-selecting of $\pi=1$ occur  {\it
before}
the post-selection of ${\cal S}$. On the other hand,
in ${\cal O}_2$  the post-selection of $\md$ seem to
take place  {\it after}
the post-selection of $\cal S$.
Nevertheless, both  observers calculated the same
probability distributions
 for  the spectrum of $A$. Probabilities are Lorentz
invariant.
However, suppose we now ask observers in  ${\cal O}_1$
and ${\cal O}_2$ to  describe the evolution of the state
of the system during a particular
measurement.
The standard interpretation,
yield two totally different descriptions.
According to the description given in frame  ${\cal
O}_1$, the  determination
of the condition $\pi_f=1$ of the $\md$,  induces a
non-local reduction
of the wave function of $\S$ {\it before}  the
condition $\psi_f$ has occurred (Figure 2).
On the other hand, a second equally valid description give by
${\cal O}_2$  is that
the determination
of $\psi_f$ occurs before, and hence
causes   a non-local collapse of the pointer {\it
before}  the  event that recorded $\pi=1$ occurred.
Obviously,  the reduction invalidates any possibility of
providing a
Lorentz covariant description in terms of wave
functions.
\medskip
\begin{center}
\leavevmode
\epsfysize=7.0cm
\epsfbox{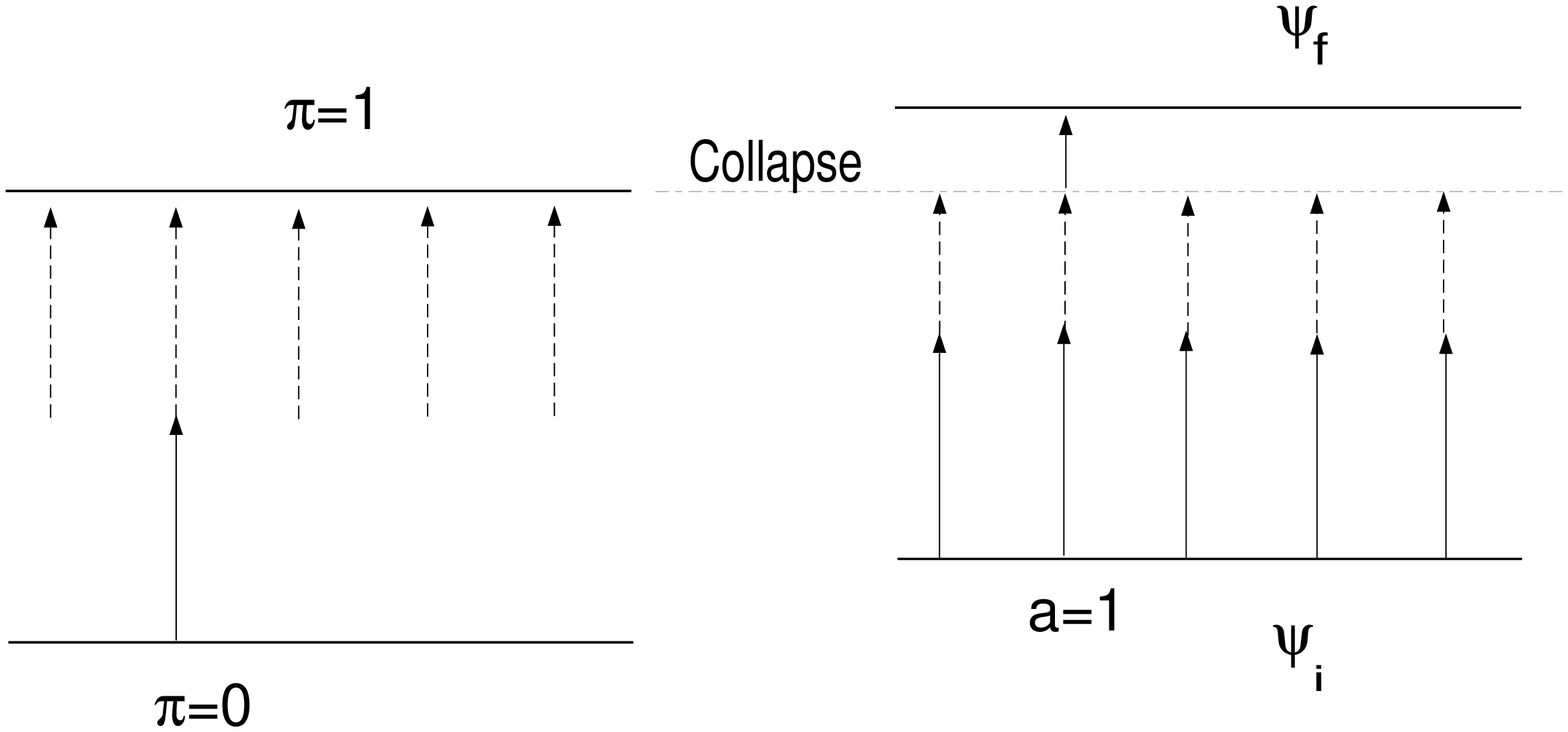}
\end{center}
\noindent
{\small
Figure 2. The evolution of the wave function in
reference frame  ${\cal O}_1$
according to the traditional interpretation. Since the
final post selection of
$\md$ and $\s$ takes place in two space-like related
locations, an
observer in ${\cal O}_1$
sees the recording o f $\pi=1$ take place  before the
final post selection of $\s$.
}
\medskip

In the two-state formulation, there is no
collapse in non of the Lorentz frames ${\cal O}_1$ or
${\cal O}_2$ described above.
In both cases we continue to describe the evolution by
using the {\it non-collapsed states}.
The schematic description given by ${\cal O}_1$ in this
case is depicted in Figure 3.
Notice that the two-state of $\s$  after the
post-selection of $\md$
is {\it still correlated} with the two-state of the
$\md$  {\it before} the post-selection. In a general
Lorentz frame the total system, $\s+\md$ is most
naturally described in terms of the multiple states
discussed
in section 2.2.
All the Lorentz frames  will use the same
multiple-state, up to the time ordering of {\it local}
conditions at space-like separated regions.
Therefore, multiple-states can provide a Lorentz
Covariant description.
\medskip
\begin{center}
\leavevmode
\epsfysize=7.5cm
\epsfbox{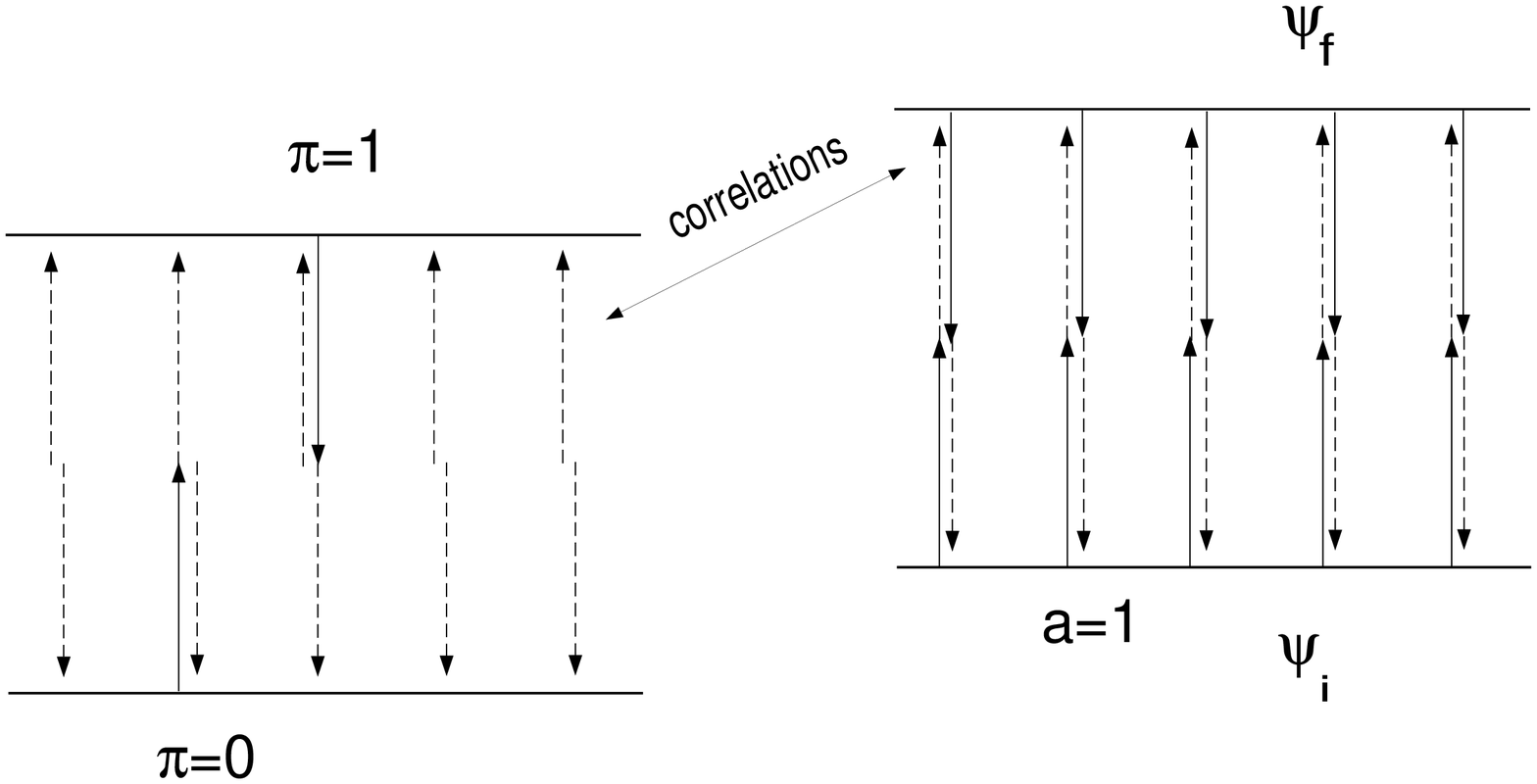}
\end{center}
\noindent
{\small
 Figure 3. The evolution of the two-state during the
measurement
in reference frame ${\cal O}_1$. There is no reduction.
Instead there are
additional time  like correlation.
}
\medskip

%In the usual description only one state is used to
describe $\cal S$ and
%therefore, one must use the collapse to be able to
calculate the probabilities
%related to intermediate measurements. In the time
symmetric formulation at
%each instant of time the
%two-state contains additional information which is
determined by
%the second
%condition. The probabilities are derived without the
mathematical
%or descriptive use of a non-local reduction.

\subsection{Measurements of non-probabilistic
observables}

In Section 2.3 we have presented a class of
complex-valued amplitude-like quantities which we have
said are non-probabilistic observables.
The  `weak values' of Hermitian operators, which can be
expressed as
$\sum_a C_a \vra(a,a)$, is a subclass of these
observables.
We shall now discuss measurements of weak values
and of other amplitude-like observables. We shall show
how  non-diagonal elements of the two-state, i.e.
$\vra(a,b)$,
which generally  can be expressed as `weak values' of
non-Hermitian operators, can be measured as well.

A consequence of the condition (\ref{strongc}) for an
accurate and hence `strong measurement', is that the
conjugate variable $q$ is  strongly fluctuating and the
coupling between S and MD (see  $H_I$ in Eq. (\ref{md}))
is large.
Therefore, any observable that does not commute with the
measured observable
 $A$ is strongly disturbed.
If we try to weaken $H_I$ by making $g_0\Delta q$ small,
we indeed disturb less
the system $\s$. However,  since $\Delta\pi$ becomes
large
we obtain a less accurate measurement of $\hat A$.
In other words, by making the location of the pointer
uncertain, we can not say if the distribution of the
results we
have obtained is due to the uncertainty $\Delta \pi$ in
the  location of
the pointer, or due to the probability distribution of
$\hat A$
which
is obtained in a ``good'' measurement. In the
  limit  ${\Delta \pi\over g_0}\to\infty, \ \
g_0\Delta q\to0$ the system S
is {\it undisturbed} at all, that is $H_I|\psi\ra\to 0
$. At first, it  may seem that
this limit is uninteresting since we can not extract any
information on the system.
However, as long as we do not set $\Delta q=0 $
identically,
we can still observe the changes in the wave function of
the pointer
while causing  the smallest disturbance we wish to the
system.
Indeed, since there is a large uncertainty in  the
location of the
pointer we shall need a large number of measurements to
find the modification of the pointer's wave function.
However, in this limit the
uncertainty is a {\it property of the measurement device
and not of the
system under observation}. In this weak interaction
limit, the evolution of the state $|\md\ra$, takes a
simple and universal form:
\beq
|\md(t) \ra = \lim_{g_0\Delta
q\to0}\la\psi_f(\s)|U|\psi_i(\s)\ra=
N(t)\exp{\Bigl(-i/\hbar\int (H_I)_w dt \Bigr)}
|\md(t=0)\ra
\label{mdt}
\eeq
For the special case
$H_I$ that corresponds to a von-Neumann coupling
(\ref{md}), this yields
\beq
\psi_{\md}(\pi,t) = N(t)\psi_{\md}(\pi-A_w, t=0).
\eeq
The initial wave function of the $\md$ is shifted by
the real part of $A_w$. The imaginary part of the weak
value can also be measured. For example, when the
initial wave function of the $\md$  is a gaussian, the
imaginary part of $A_w$ affects
the `velocity' of the pointer, which  in our case is
represented by the  $q$-coordinate.
Notice that the wave function of {\it all} the
measurement devices in the
ensemble are modified in the {\it same} way. In
principle this can be
confirmed by projecting
the final state of the pointer on the computed
projection operator $|\md(t)\ra
\la \md (t)|$. In the usual case, one determines the
final state  of the pointer in $\pi$-space.
Therefore, an ensemble of measurement devices is needed
only to eliminate the (known) uncertainty in $\pi$.

We now consider an alternative measurement set up which
can be used to measure the two-amplitude $\vra(a,b)$.
Since $\vra(a,b)=(\vr_{ab})_w/\la\vr_{ab}, \vr_{ab}\ra$,
we need actually to measure weakly
the non-Hermitian operator $A_{ab}\equiv  (\la b|a\ra)
|a\ra\la b|$.
This can be achieved by the following modification of
the usual procedure.
We add a {\it third} device, which is  large spin $L=N$,
and pre and post-select the rare states  $L_z=N$ and
$L_x=N$, respectively. At an intermediate time we set
the interaction
\beq
H_I = {g(t)\over \sqrt2 N} q (A^\dagger_{ab}L_+  +
A_{ab} L_-)
\label{hi}
\eeq
We find that
the evolution of the $\md$ is given by
\beq
\psi_{\md}(\pi,t) = C(t)\psi_{\md}(\pi-\vra(a,b), t=0)
+O(g_0\Delta q/N)
\eeq
The idea of  this procedure is to achieve an effective
coupling with a non-Hermitian operator. Although the
total interaction is Hermitian,  this specific  pre- and
post-selection of the large spin,  makes the
contribution of the term with $L_-$ negligible, while
leaving the second terms as the main contribution.
When the correction $O(g_0\Delta q/N)$ is negligible, we
obtain a measurement
of the two-amplitude $\vra(a,b)$. Note that we need
either a small $g_0\Delta q$
or a large $N$. In the first case our coupling yields a
`weak' measurement
of $A_{ab}$. However in the case of large $N$ we can
regard our coupling
as an ordinary measurement, i.e. for every given finite
accuracy $\Delta \pi$
of our measuring device, we use a sufficiently large $N$
such that we
{\it always} measure $\vra(a,b)$. Of course, in the
latter case we need to work harder in order to prepare
our ensemble.
The ``large-$N$ limit'' can of course be used in
measuring weak values of Hermitian operators as well.

The common property of the two limits is that in both
cases we can regard the
effect of the interaction (\ref{hi}) on the observed
system $\s$ as
very small, i.e. $H_I|\psi\ra \psi_{\s} \sim O(\Delta
q/N)$.
Therefore, in the limit, the wave function of the system
is unmodified.

\subsection{The intermediate regime: mixing of
probabilities and weak values}

In the previews two sections we have considered
measurements that according to the strength of the
coupling, could be classified either as  strong or as
weak measurements. In the first case, the results are
described by a probability distribution, while in the
second case, they are interpreted as a measure of
essentially  non-probabilistic two-state amplitudes.
What happens when the strength of the coupling
correspond to some intermediate regime and  the accuracy
of the measurement is not sufficient for a strong
measurement and  too small to be regarded as a weak
measurement?

We shall now  show ,that at least in some cases,  in the
intermediate regime,  we  measure observables which are
expressed by a  mixing of probabilities and
amplitude-like quantities.
Suppose that the system under observation is pre- and
post-selected to a two-state
$\vr_\s=|\psi_{in}\ra\la\psi_{out}|$, and that the
measurement device in is initially in the  state
$|\md(0)\ra$.
Then, restoring the corrections previously omitted in
equation (\ref{mdt}), the final state of the measurement
device is  given by
\beq
|\md(t) \ra = \biggl[ \exp(-ig_0qA_w) +
\sum_{n=2}^\infty {(-ig_0 q)^n\over n!} \Delta A_w^n
\biggr] \ |\md(0)\ra,
\label{mde}
\eeq
where $\Delta A_w^n \equiv(A^n)_w - (A_w)^n$.
The `weak' approximation requires that the `evolution
operator' above is given only by the exponential term.
If the sum above is dominated by the first term, then a
sufficient condition for a weak measurement is that
$g_0^2 |\Delta A_w^2| \Delta q^2<<1$.
Now suppose that this condition is not satisfied for our
given two-state $\vr_s$, but  we can still find a
decomposition in terms of normalized two-states $\vr_k$
\beq
\vr_s =  \sum_k a_k \vr_k,
\label{components}
\eeq
such that each of the component $\vr_k$ satisfies
\beq
 g_0^2 |\Delta(A_w^2)_k| \Delta q^2<<1,
\eeq
Here,  $(A_w)_k=\tr(A\vr_k)/\tr\vr_k$ is  the weak value
of $A$ with respect to the $k$ component of the
two-state. Although for this {\it given} coupling
strength $g_0^2\Delta q^2$, the `weak uncertainty'
$\Delta A_w^2$ for the two-state $\vr_s$ is not
sufficiently small,
in each of the components $\vr_k$ the  `weakness'
condition is satisfied.

Pictorially we can clarify the meaning of this
condition as follows. In order to obtain a weak
measurement we need that the uncertainty in the
measurement will be larger then the given uncertainty of
the observable. If $A$ is distributed in several
disconnected areas, say  $A\in \Delta_k$ $, \ \
k=1,..l$, then generally the total uncertainty could be
larger then the uncertainty in each of the component,
i.e. $\Delta A>> \max_k(\Delta_k)$.
Due to the existence of these two scales,  it is quite
possible, that while the accuracy of the measurement is
too high to yield a weak measurement of $A$ for the
total two-state, (since it can differentiate between the
different branches $\vr_k$ of $\vr$), it is sufficiently
large  for each of the components with  smaller
uncertainty $\Delta_k$.

We can now rewrite  equation (\ref{mde}) as
$$
|\md(t) \ra \simeq N\sum_k a_k \biggl[
\exp\Bigl(-ig_0q(A_w)_k\Bigr) +  {(-ig_0 q)^2\over n!}
\Delta(A_w^2)_k \biggr] |\md(0)\ra
$$
\beq
\simeq N\sum_k a_k  \exp\bigl(-ig_0q(A_w)_k\bigr) \
|\md(0)\ra
\eeq
or,
\beq
|\md(t)\ra \simeq N\sum_k a_k |\psi_\md(\pi-
(A_w)_k))\ra
\eeq
Since at each of the measurements one of the components
is selected with probability $|a_k|^2$, this measurement
determines the `averaged weak value'
\beq
{1\over\sum_k|a_k|^2}\sum_k |a_k|^2 (A_w)_k
\label{intermediate}
\eeq
This mixed average, can be contrasted  with the purely
amplitude-like weak value which by equations
(\ref{transformation}) and  (\ref{components}) is given
by
\beq
{1\over\sum_ka_k}\sum_k a_k (A_w)_k
\eeq

To exemplify this interesting case, consider the system
$\s$ to be a large spin with a maximal value $L=N$.
Let the system be pre-selected in the state \hfill\break
$|\psi_1\ra= a'|L_x=N\ra + b'|L_x = -N\ra$, and
post-selected in the state
$|\psi_2\ra = |L_y=N\ra$.
Thus the two-state is given (for $H=0$) by
\beq
\vr = a\vr_{+} + b\vr_{-}
\eeq
where $\vr_{\pm } = |L_x=\pm N\ra \la L_y=N|/\la
L_y=N|L_x=\pm N\ra$ are normalized two-states,
$a=\la L_y=N|L_x=N\ra a'$, and $b=\la L_y=N|L_x=-N\ra
b'$.
We choose the operator to be observed as
\beq
A={1\over\sqrt2}(L_x+L_y).
\eeq
The weak value of $A$ is
\beq
A_w = {1\over\sqrt2}N\biggl(1+ {(L_x)_w\over
N}\biggr)\sim {1\over\sqrt2} N
\label{axyw}
\eeq
In the two branches $\vr_{\pm}$ we have
\beq
A_{w+} =\sqrt2 N, \ \ \ \ \ A_{w-} = 0
\eeq
The `weak uncertainty' of $A$ in the two-state $\vr$ is
\beq
\Delta A_w = {1\over2}\biggl(N^2\Bigl[1-{(L_x)_w\over
N}\Bigr] + i(L_z)_w\biggr)
\simeq{1\over2}(N^2 +iN),
\eeq
while in the two branches
\beq
\Delta A_{w\pm} = {i\over2}(L_z)_{w\pm}\simeq {i\over2}N
\eeq
Therefore, for a sufficiently large $N$,  we have two
scales.
For $g^2_0\Delta q^2 << 1/N^2$ we shall obtain the weak
value (\ref{axyw}),
but in the range $1/N^2<<g^2_0\Delta q^2<<1/N$ we shall
measure the mixed quantity
\beq
{1\over |a|^2+|b|^2} \biggl(|a|^2 A_{w+} +|b|^2
A_{w-}\biggr).
\eeq

\vfill\eject

%%%%%%%%%%%%%%%%%%%%%%%%%%%%%%%%%%%%%%%%%%%%%%%%%%%%%%%%
%%%%%%%%%%%%%%

\section{Conceptual implications}

In this section we re-examine some possible
implications
of the two-state formalism to well known conceptual
problems in quantum mechanics.
We shall suggest that by replacing the wave function by
the two-state
as the fundamental object,  the problem of non-local
reduction can be avoided.

\subsection{The EPR experiment}

To set notations, suppose an observer in the
`rest frame' ${\cal O}$ prepares at $t<-T$ two particles
with an internal  spin $1/2$ degree of freedom, in a
singlet state. At $t=-T$ the initial state is
\beq
|\psi(-T)\ra \ = \
{1\over\sqrt2}( |\up\ra_1 |\down\ra_2 -
|\down\ra_1|\up\ra_2 )
\label{singlet}
\eeq
The indices $1,2$ stand for the spatial location of the
particles at $x_1$ and $x_2=x_1+L$, respectively. The
distance $L$ between the particles can be  arbitrarily
large.
Suppose that at $t=+T$, an observer measures
$\sigma_1=\hat n_1\cdot\vec \sigma_1$ and at
$t=T+\epsilon$ (the spin of 1 in the $\hat n_1$
direction) and
another observer measures $\sigma_2\hat
n_2\cdot\vec\sigma_2$. The usual way to describe  the
evolution of the state is to say that the wave function
(\ref{singlet})
should be reduced
according to the result of the first measurement.
At $t=T+\epsilon$, the correlation between the particles
is already
washed out and the wave function of particle $2$ is
given by $\la \sigma_1|\psi\ra$.
This description involves a non-local reduction of
$|\psi\ra$ which is clearly
not covariant. An observer in  a moving frame ${\cal
O}'$
observes the measurement at site $2$ take place {\it
first}, hence
he will reduce $|\psi\ra$ according to the observed
value of $\sigma_2$.
{}From a practical point of view this discrepancy is not a
problem. Probabilities are Lorentz invariant quantities.

However, from the conceptual point of view,
it presents a deep difficulty. Can we attribute any
reality to
the wave function if
two observers $\o$ and $\o'$ describe the evolution of
the system in
two totally  different ways?

To this well known criticism we would like now to add
the following.
We can define or relate to a ``physical collapse'' the
following  operational meaning.
Consider a  measurement described by the von-Neumann
coupling
\beq
H_I \ = \
g_0\Bigl(\delta(t-\xi) - \delta(t+\xi)\Bigr)q\sigma_z
\label{meas-collapse}
\eeq
We imagine the measuring apparatus as another  quantum
system and read
of the result of the measurement by coupling it to a
macroscopic large
system (`the environment') only after $t=\xi$.
Suppose that  the measurement
device was prepared at $t<-\xi$ and was left undisturbed
at
$t\in(\xi,-\xi)$. Then,  the final reading at $t>\xi$
yields the value
$\delta\pi = \pi_f-\pi_i=g_0\Bigl(\sigma_z(t=\xi)-\sigma_z(t=-\xi)
\Bigr)$.
If the evolution of the spin (and the measuring device)
in the time interval
$t\in(\xi,-\xi)$ was undisturbed, then we can predicted
with probability 1
that $\delta\pi=0$.
However, if
at $t=0$ the value of say $\sigma_x$ was measured by
some other device, or if some
other interaction took place,
then  the  evolution in this time interval would be
disturbed
and the
result would generally by given by $\delta\pi\ne0$ !
Therefore, we have a physical criteria to identify a
reduction of the state.

Returning to the EPR experiment, let us assume that $\o$
measured
$\sigma_{1z}$ and then uses our apparatus
(\ref{meas-collapse})
to search some discontinuity in the evolution of
$\sigma_2$. Clearly, he will find
$\delta\pi=g_0(\sigma_{2z}(t=T+\xi)-\sigma_{2z}(t=T-\xi)
)=0$  {\it always}!.
Similarly the observer in the frame $\o'$
may confirm that the collapse for the spin $\sigma_2$
did not take place on his hypersurface
of simultaneity. Although this argument does not role
out the possibility of a non-local reduction, it shows
that while we can operationally identify a local
reduction, we cannot by the same measurement identify a
non-local reduction.
This again suggests that
non-local
reduction of the
wave function may not be a real physical process.
Nevertheless it is possible that there exists a {\it
local}
physical process of reduction of the wave function.

If we  assume that a non-local reduction is not a
physical
process. How should we then describe the state of the
system after  observation,
and how can we calculate and find
the (non-local) correlations
in the
EPR experiment?

Let us now examine the EPR experiment in the context of
the two-state formulation. The state of the system is
fully described only
when two  conditions are determined for both particles.
The first
 condition, $|\psi_1\ra$ is in this case a singlet
state.
The second  condition is provided by the values of
$\sigma_1$ and $\sigma_2$, i.e. by $|\psi_2\ra =
|\sigma_1\ra\otimes|\sigma_2\ra$.
Hence, in  the case $H_1=H_2=0$, the normalized
two-state that corresponds to
the EPR experiment is given by
\beq
\vr_{EPR} \ = \
{1\over2}\Bigl( |\up_z\ra_1\otimes|\down_z\ra_2 -
|\down_z\ra_1
\otimes|\up_z\ra_2 \Bigr)\Bigl(\la \sigma_1|\otimes\la
\sigma_2|\Bigr) .
\label{epr}
\eeq
The EPR two-state is Lorentz covariant since it is
completely determined
local  conditions, which are a result of local
observations of the
spin.
To retain the usual probabilistic information
consider for example the case
we found $\sigma_{1z}=1$. The probability to
measure $\sigma_{2\hat n}=\pm{1}$, for the spin of
particle 2
in the direction $\hat n$ is obtained as a conditional
 probability which is derived
from the two-states $\vr(\sigma_{2\hat
n}={1})\equiv\vr(\up_{2\hat n})$
 and
$\vr(\sigma_{2\hat n}=-{1})\equiv\vr(\down_{2\hat n})$.
The latter correspond to the two (only) possible final
 conditions obtained by an observation of the
 spin of particle $2$ in the
$\hat n$ direction. We first calculate $\rho_{in} =
\vr\vr^\dagger/\tr(\vr\vr^\dagger)$
and $\rho_{out}=\vr^\dagger\vr/\tr(\vr^\dagger\vr)$.
 The probability is then
expressed by
\beq
\prob(\up_{2z}) = {\la \rho_{out}(\up_{2z}), \rho_{in}
\ra
\over
\la \rho_{out}(\up_{2z}), \rho_{in} \ra + \la
\rho_{out}(\down_{2z}), \rho_{in} \ra }
\label{pepr}
\eeq
For $\hat n = \hat z$ we can form only the two-state
$\vr(\down_{2z})$,
while for $|S_{2z}\ra=|\up_{2z}\ra$ we {\it do not have}
a corresponding two-state
$\vr(\up_{2z})\in \h_{phys}$.
In this case  $\la\psi_2|\psi_1\ra=0$ and we can not
form a
normalized $(\tr\vr=1)$ two-state.
Since we have only one possible two-state, the
conditional probability equals $1$.

To summarize, our description of an EPR experiment by
means two-state in equation (\ref{epr}) is Lorentz
covariant.
There is no element of non-local reduction since the
information on the final results is coded in the final
{\it local}  conditions.
Finally, probability distributions my be restored by
constructing conditional
probabilities as in equation (\ref{pepr}), i.e. by
comparing different two-state ensembles.

\subsection{Repeated measurements without reduction}

In the usual description of repeated measurements, the
state of the
observed system $\s$ is viewed as changing
discontinuously  after each
observation. For example, consider successive
measurement
of $x,p,x,..$ ,or any other two non-commuting
observables.
These discontinuities generally correspond to  non-local
reductions of the wave function.

We now argue  that in the two-state  formulation, the
evolution of
 the system $\s$ {\it is continuous}
and the only (possible) local-reduction takes place
at the measurement device.
Let us consider a system $\s$
and two measurement devices $\md_1$ and $\md_2$,
with the initial conditions
$|\psi_1\ra  = |\pi_1=0\ra\otimes|\pi_2=0\ra\otimes\sum
C_n|A=n\ra $
at $t=0$.
The interaction Hamiltonian given by
\beq
H_I = g_0\Bigl(
    \delta(t-t_1) q_1A + \delta(t-t_2) q_2 B \Bigr) .
\eeq
At $t=t_1$, $\md_1$ interacts with $\s$ and at
$t=T_1=t_1+\Delta$ the result
$\pi_1=a$ is recorded on some macroscopic body.
Latter, at $t=t_2$, $\md_2$ interacts too with $\s$, and
the result $\pi_2=b$ is
recorded on a macroscopic body at time
$t=T_2=t_2+\Delta$.
The time interval, $\Delta$,
between the interaction and the final reading of $\pi$,
due to some
coupling to an `external' environment, is finite but
otherwise can be arbitrary.
A schematic
evolution of the system in the `forward' and `backward'
directions of time
is  represented in Figure 4.
As long as the final state of $\s$ is unknown we can not
fully
determine the two-state of the system.
The probability distribution for finding $\pi_1=a$ and
$\pi_2=b$ depends on the final  condition $|\psi_f\ra$
(obtained by post selection)
of $\s$ at $t=T$. Therefore, if the observations by
$\md_1$ and $\md_2$
where performed only on a pre-selected ensemble we must
average
over all final possible states, i.e. consider
conditional probabilities of
different two-time ensembles.

For example let us consider the case of only one (known)
measurement.
Suppose that at some time at the future a some Hermitian
operator $\hat K$
with eigenfunctions $|\psi_k\ra$ is measured.
Therefore {\it one} of the two-states $\vr_k$ has been
determined but is unknown to us. Therefore, the
probability $\prob_I(a)$ to measure $a$ is given by
\beq
\prob_I(a) = \sum_k \prob(a; \vr_k)\prob(\psi_k),
\label{sum}
\eeq
where $\prob(a;\vr_k) $ and $\prob(\psi_k; \vr_k)$
are the probability to find $A=a$ (
given that the final state is $\psi_k$), and the
probability to find $\psi_k$, respectively. A
straightforward  substitution yields $\prob_I(a) = |\la
a |\psi(initial)\ra|^2$ as expected. Notice that this
result does not depend on what observable is actually
measured in the future.
In a similar way one can reconstruct the probability to
find $B=b$ at the second measurement. Therefore, as
before all the usual probabilistic information may be
obtained.
%\medskip
\begin{center}
\leavevmode
\epsfysize=10.5cm
\epsfbox{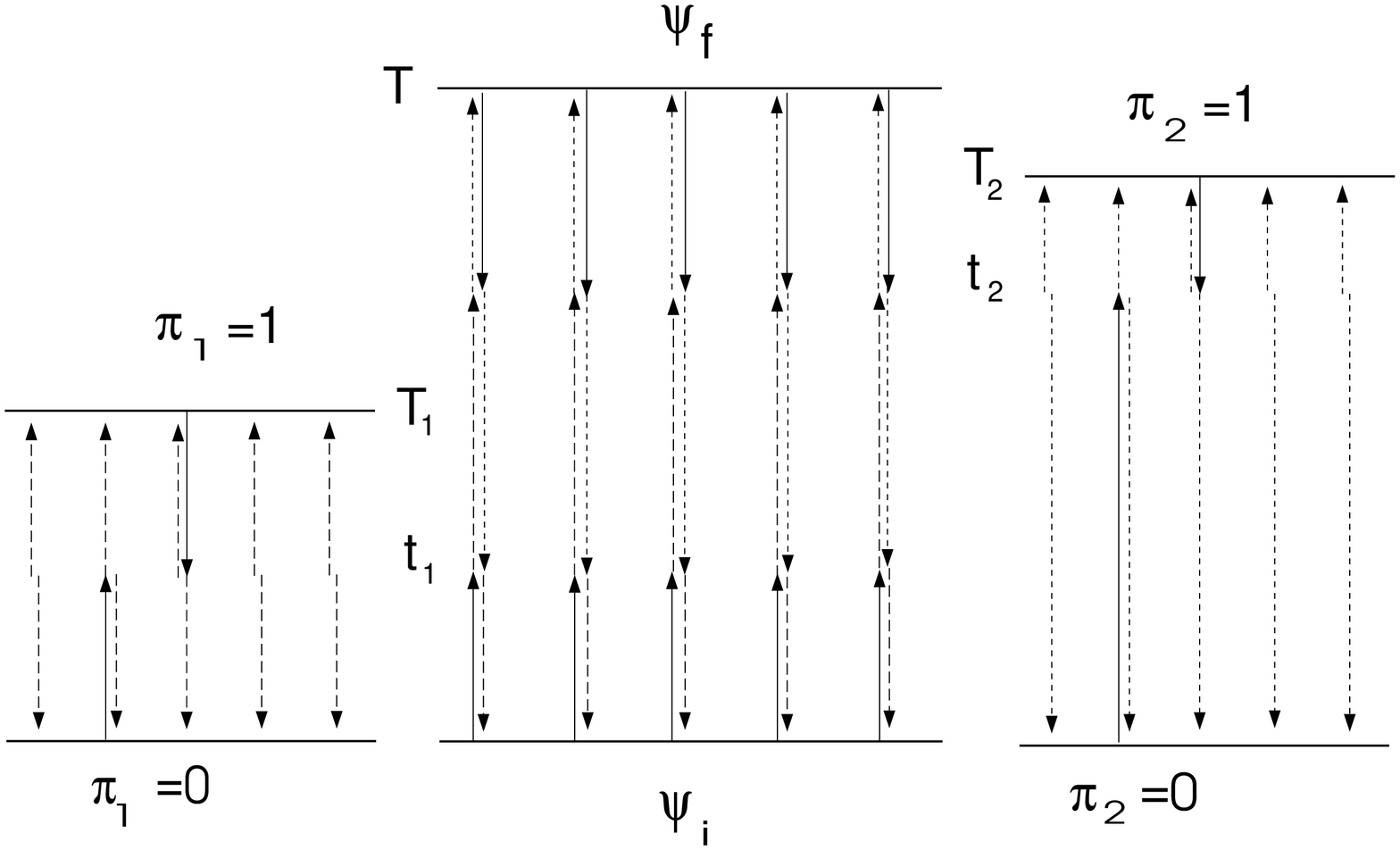}
\end{center}
\noindent
{\small
Figure 4:  Pictorial description of the two-state
(\ref{succ}) of a system under two
successive observations, in the special case
of a successive measurement of the same observable with
the result
$A=1$. At $t=t_1$, $\md_1$ (on the left) interacts with
$\s$ and at
$t=T_1$ $\md_1$ is post-selected to a final state
with $\pi_1=1$.
At $t=t_2$, a second measuring device $\md_2$ (on the
right)  interacts with $\s$, and
post-selected to a final state with $\pi_2=b$ at
$t=T_2$. Finally at $t=T$ the system is post-selected to
a  final state $\psi_f$.
Correlations between $\md_1$ and $\s$ are denoted by
dashed lines, and with $\md_2$ by dotted lines.
The two measuring devices must yield the same result
with probability one because for any other result
$\tr\vr = 0$.
}
\medskip

Only in the special case, when the same observable is
measured twice, i.e. $A=B$,
we find that {\it for every} final state we must have
$\pi_1=\pi_2=a$.
When this condition is not satisfied we find that for
every initial
and final state of $\s$, the initial state of the total
system
can not evolve to the final state, i.e., $\tr\vr =\la
\psi_1|\psi_2\ra =0$.
Therefore, in this special case, the two measurements
must yield the
same result with probability one.

Therefore, let us assume that the final state of $\s$
has been determined  and  consider the evolution of the
two-state  in the case of repeated measurements.
Since the  two-state is determined only by the local
conditions  the state of $\s$ is not reduced after
the coupling with $\md_1$ and $\md_2$.
However we do pay a prize for avoiding the reduction,
which is the necessity of
including in our description of the total
system  time-like correlations.
As depicted in Figure 4, the forward evolving state of
$\md_1$ at
$t\in(t_1,T_1)$ remains correlated to the
state of $\s$ at $t>T_1$. Similarly, the forward
evolving state of
$\md_2$ at $t\in(t_2,T_2)$ becomes correlated with $\s$
and hence
also with $\md_1$. These time-like correlations are
natural from
the point of view of our formalism.
The  multiple-state of the total system is  generally
given by:
\beq
\vr(t_1,t_2,t_3) = \sum
C_{ijklmn}(t_1,t_2,t_3)\vr_{\md1ij}(t_1)\otimes
\vr_{\md2 kl}(t_2)\otimes\vr_{\s mn}(t_3),
\label{succ}
\eeq
 where $t_1\in(0,T_1), \ \ t_2\in(0,T_2)$ and $ \
t_3\in(0,T)$.

%\vfill\eject

%%%%%%%%%%%%%%%%%%%%%%%%%%%%%%%%%%%%%%%%%%%%%%%%%%%%%%%%
%%%%%%%%%%%%

\section{Discussion}

The first part of this article was devoted to a formal
construction of the two-state formalism. We have seen
that this formalism incorporate in a natural way two
basic
classes of observables.  Probabilistic observables which
arise whenever a
system is observed by means of a (strong) demolition
experiment, and complex amplitude-like observables which
are
measured in any non-demolition (weak) experiment.
These amplitude-like observables include as a subclass,
the weak values of hermitian operators. The second class
of observables is also
 related to the  recent proposal for a  ``measurement of
the wave function''\cite{schrodinger-wave}. To see the
connection,  consider a system, with $H=0$,
which is
  pre- and post-selected in the same wave function
$\psi(x)$. In such circumstances, the weak value of the
projection operator, $\int_\Delta dx |x\ra\la x|$,
is given by the average value of $|\psi(x)|^2$ in the
domain $x\in\Delta$.
However, by Eq. (\ref{vraxx}), $|\psi(x)|^2=\vra(x,x)$,
i.e. it is the diagonal element of the two-amplitude.
Therefore, the same quantity, which is being measured in
Ref. \cite{schrodinger-wave} by means of an adiabatic
process, can be obtained also by a weak measurement.
A way to measure the two-state is suggested also in Ref.
\cite{protective}.
We have also discovered that  in the intermediate regime
between strong and weak measurements, there can exist an
amusing mixing of probabilities and weak values.

We have shown that the two-state formalism has also
conceptual advantages.
By recasting measurement theory in terms of two states
as elementary objects, it seems that
 we came
closer to formulating a sensible consistent
interpretation
of the measurement process.
We did not eliminate completely the element of
reduction, but instead we used conditions.
However, by avoiding the non-local reduction, we opened
the possibility
of incorporating  consistent local physics.
Another possibility is that there is no local physical
process of reduction, and that the solution may be found
by handling the  conditions of a closed system in a
dynamical way. In this program one would like to
eliminate some `special'   initial and  final conditions
which yield a consistency of the total history.

\vfill\eject

\section{Appendix}

In this appendix we shall show that non-generic two-
states can describe sub-systems. For further discussion
see ref.  \cite{complete,trace}). Consider two
non-interacting systems
$\tilde\s$ and $\s$ that are pre- and post selected in
the following states:
\beq
|\Psi_{in}(t=0)\ra = \sum_{nm}a_{nm}
\tilde{|\phi_n\ra}\otimes|\psi_m\ra
\eeq
and
\beq
|\Psi_{out}(t=T)\ra  = \sum_{ij} b_{ij}
\tilde{|\phi_i\ra}\otimes|\xi_j\ra
\eeq
$\tilde{\lbrace|\phi_n\ra\rbrace}$ is an orthonormal
basis of the Hilbert space $\tilde\h$ of $\tilde\s$,
($\tilde{\la\phi_n}|\tilde{\phi_m\ra}=\delta_{nm}$).
$\lbrace|\psi_i\ra\rbrace$ and $\lbrace|\xi_j\ra\rbrace$
are two orthonormal basis of the Hilbert space $\h$ of
$\s$ but with the property $\la\psi_i|\xi_j\ra\ne0$ for
all $i,j$.

The total system is described by the generic two-state
$\vr_{total}=|\Psi_{in}\ra\la\Psi_{out}|$. The
probability of measure the eigenvalue $\lambda$ of some
general operator acting in $\tilde\h\otimes\h$
is
\beq
\prob(\lambda)= N |\tr(\pi_\lambda \vr_{total})|^2,
\label{lambda}
\eeq
where $N$ is the normalization, and
$\pi_\lambda=|\lambda\ra\la\lambda|$.
Now suppose we are interested in measuring observables
that are related
only to $\s$, i.e. an Hermitian operators that acts in
$\h$.
In this case, equation (\ref{lambda}) can be replaced
by
\beq
\prob(\lambda) =  N |\tr(\pi_\lambda \vr_{eff})|^2
\eeq
where
\beq
\vr_{eff} = \sum c_{ij} |\psi_i\ra\la\xi_j|, \ \ \ \
c_{ij}=\sum_n a_{ni}b^*_{nj}.
\eeq
is the reduced effective two-state.
$\vr_{eff}$ is a non-generic two-state.
Generic two-states correspond to a complete
specification of the initial and final conditions for
the system. When the conditions are determined only
``partially'' the system is initially and finally in a
mixed state. In the context of our formalism this can
be interpreted as a situation with
correlations between the initial and final  conditions.

\vspace{.3in}
\noindent
{\bf Acknowledgment}\\

We would like to thank to Lev Vaidman for useful
comments.

\vfill \eject

\end{document}